\documentclass[12pt]{iopart}
\usepackage{iopams}  
\usepackage{graphicx}
\usepackage{epstopdf}
\usepackage{cite}
\begin{document}


\title[Structural and electronic properties of graphene/Al/Ni(111)]{Structural and electronic properties of the graphene/Al/Ni(111) intercalation-like system}

\author{E. N. Voloshina,$^{1}$ A. Generalov,$^{2,3}$ M. Weser,$^{2}$ S. B\"ottcher,$^{2}$\\ K. Horn,$^{2}$ and Yu.\,S.\,Dedkov$^{2,3}$}
\address{$^1$Physikalische und Theoretische Chemie, Freie Universit\"at Berlin, 14195 Berlin, Germany}
\address{$^2$Fritz-Haber-Institut der Max-Planck-Gesellschaft, 14195 Berlin, Germany}
\address{$^3$Institut f\"ur Festk\"orperphysik, Technische Universit\"at Dresden, 01062 Dresden, Germany}
\ead{elena.voloshina@fu-berlin.de}
\ead{yuriy.dedkov@googlemail.com}

\begin{abstract}
Decoupling of the graphene layer from the ferromagnetic substrate via intercalation of $sp$ metal has recently been proposed as an effective way to realize single-layer graphene-based spin-filter. Here, the structural and electronic properties of the prototype system, graphene/Al/Ni(111), are investigated via combination of electron diffraction and spectroscopic methods. These studies are accompanied by state-of-the-art electronic structure calculations. The properties of this prospective Al-intercalation-like system and its possible implementations in future graphene-based devices are discussed. 
\end{abstract}

\pacs{61.05.cj, 68.65.Pq, 73.20.-r, 73.22.Pr, 78.70.Dm, 79.60.-i}

\maketitle

\section{Introduction}\label{introduction}

Graphene, an one-atom thick carbon layer~\cite{Geim:2007a,Geim:2009,Geim:2011,Novoselov:2011}, is the thinnest conductive material in nature. Presently, many applications on the basis of graphene were proposed and/or realized, which are using its unique electronic structure in the vicinity of the Fermi level, $E_F$ [linear dependence of energy ($E$) on momentum of electron ($k$), $E=\hbar v_F k$, where $v_F\approx1\times10^6$\,m/s is the Fermi velocity, which plays the role of the ``speed of light'' for the mass-less Dirac fermions in graphene]. Among them are precision gas sensors~\cite{Schedin:2007}, transistors with a cutoff frequency of 100\,GHz~\cite{Lin:2010}, flexible touch-screens~\cite{Bae:2010}, integrated circuits~\cite{Sprinkle:2010,Lin:2011}, and many others~\cite{Geim:2009,Novoselov:2011}. One of the promising applications of graphene is its use as an effective spin-filtering material~\cite{Karpan:2007,Karpan:2008,Cho:2011}. In such graphene-based spin-filter one or several layers of graphene are sandwiched between two layers of a ferromagnetic (FM) material [current perpendicular to plane (CPP) geometry]. As ferromagnetic electrodes the close packed surfaces Ni(111)~\cite{Dedkov:2010a}, Co(0001)~\cite{Eom:2009}, or Fe(111)/Ni(111)~\cite{Dedkov:2008b,Weser:2011} interface have to be used. Here, the spin-filtering effect originates from the fact that the projection of the Fermi surface of FM on to the close-packed (111) plane overlaps with the Fermi surface of graphene (single points at the $K$ points of the hexagonal Brillouin zone of graphene) only for electrons of one kind of spin, for spin-down electrons~\cite{Karpan:2007,Karpan:2008}. A similar effect was predicted for the spin-filtering device where graphene sheet connects two ferromagnetic electrodes made from Ni~\cite{Cho:2011}. 

In the FM/graphene/FM sandwich-like structures graphene strongly interacts with the ferromagnetic layers~\cite{Weser:2011,Bertoni:2004,Karpan:2007,Karpan:2008} preventing effective spin-filtering in such a way that two spin-channels start to participate in the transport across the graphene layer. The effect of strong interaction between graphene and FM was confirmed by means of angle-resolved photoelectron spectroscopy (ARPES) and near-edge x-ray absorption spectroscopy (NEXAFS) and manifests itself in the strong hybridization of the graphene\,$\pi$ and Ni\,$3d$ states in the vicinity of $E_F$~\cite{Dedkov:2008a,Gruneis:2008,Weser:2010,Rusz:2010,Dedkov:2010a}. There are several ways to overcome this problem: (i) use of the several graphene layers, where the upper layers are decoupled from the bottom graphene layer, which is in contact with FM, and the Dirac cone is restored allowing effective spin-filtering~\cite{Karpan:2007,Karpan:2008}, (ii) application as the upper electrode the half-metallic ferromagnetic material (such material has electrons of only one kind of spin at $E_F$~\cite{Katsnelson:2008}), which is electronically decoupled from the graphene layer~\cite{Dedkov:2011a}, and/or (iii) intercalation of noble metals in the space between graphene and a ferromagnetic electrode~\cite{Shikin:1998,Shikin:2000,Dedkov:2001,Haberer:2010} that leads to the weakening of the interaction of the graphene layer and substrate with the corresponding restoring of the Dirac cone. Whereas the first two cases were not studied in detail or only sporadic studies of the electronic structure of the system can be found in the literature, the effect of intercalation of noble metals on the electronic structure of the graphene layer on metallic substrates was studied in more detail~\cite{Shikin:1998,Shikin:2000,Dedkov:2001,Haberer:2010,Varykhalov:2008,Varykhalov:2010a,SanchezBarriga:2010,Enderlein:2010}. These studies performed by means of ARPES and supported by theoretical calculations~\cite{Giovannetti:2008,Khomyakov:2009} demonstrate that a clear hybridization between the graphene\,$\pi$ and $d$ states of noble metals exists at the graphene/noble-metal interface~\cite{Varykhalov:2008,Varykhalov:2010a}. Such hybridization interaction can lead (due to the strong spin-orbit interaction~\cite{Takeuchi:1989,Wilkin:1992,Takahashi:2008}) to an effective spin-scattering at the interface between noble metal and graphene. One of the possibilities to avoid such undesirable effect is the substitution of the $d$-electron noble metals by a simple $sp$ metal, which can lead to the restoring of the Dirac cone in the vicinity of $E_F$ in contrast to the graphene/FM case, where it is destroyed by the strong interaction between graphene and the FM substrate. The effect of spin-scattering is thus avoided at the interface between graphene and a $sp$ metal.

Here, we investigate the geometry and electronic properties of the graphene/Al/ Ni(111) intercalation-like system by means of low-energy electron diffraction (LEED) and a combination of spectroscopic methods [NEXAFS, core-level photoelectron spectroscopy (CL-PES), ARPES] and state-of-the-art band structure calculations. It was found that aluminum, which has electronic configuration $3s^23p^1$, effectively decouples graphene from the substrate that Dirac cone is fully restored in the vicinity of $E_F$ contrary to the graphene/Ni(111) system. In contrast to the noble-metal graphene-based intercalation-like systems, in the present case there is no strong hybridization between graphene\,$\pi$ and $d$ states that will lead to decreasing of the effect of spin-dependent scattering at the interface. The perspectives of application of this interface as well as possibility to prepare fully epitaxial Al-oxide based system, graphene/AlO$_x$/Ni(111), are discussed. Such AlO$_x$-based intercalation-like system can be used for the effective injection of spin-polarized electrons via tunneling barrier from FM Ni into a graphene layer allowing minimization of the  interface scattering~\cite{Tombros:2007,Yang:2011}.

\section{Computational and experimental details}\label{experimental}

\textit{Calculations.} The DFT calculations were carried out using the projector augmented wave method~\cite{Blochl:1994}, a plane wave basis set and the generalized gradient approximation as parameterized by Perdew \textit{et al.} (PBE)~\cite{Perdew:1996}, as implemented in the VASP program~\cite{Kresse:1994}. The plane wave kinetic energy cutoff was set to $500$\,eV. The long-range van der Waals interactions were accounted for by means of a semiempirical DFT-D2 approach proposed by Grimme~\cite{Grimme:2004,Grimme:2006,Grimme:2010}. In the total energy calculations and during the structural relaxation (the positions of the carbon atoms as well as those of Al and the top two layers of Ni are optimized) the $k$-meshes for sampling of the supercell Brillouin zone were chosen to be as dense as $24\times 24$ and $12\times 12$, respectively, when folded up to the simple graphene unit cell. The studied system is modeled using a supercell consisting of 74 atoms. It has $(2\times2)$ lateral periodicity and contains 13 layers of Ni atoms (4 atoms per layer) with one Al layer (3 atoms each) and a graphene sheet (8 atoms per layer) adsorbed on both sides of the slab. Metallic slab replicas are separated by ca. $24$\,\AA\ in the surface normal direction, leading to an effective vacuum region of about $17$\,\AA.

The scanning tunneling microscopy (STM) images are calculated using the Tersoff-Hamann formalism~\cite{Tersoff:1985}, which states that the tunneling current in an STM experiment is proportional to the local density of states (LDOS) integrated from the Fermi level to the bias. In the visualization software Hive~\cite{VanPoucke:2008} it is implemented in its most basic formulation, approximating the STM tip by an infinitely small point source. The integrated LDOS is calculated as $\bar{\rho}(\mathbf{r},\varepsilon)\propto\int_\varepsilon^{E_F}\rho(\mathbf{r},\varepsilon')d\varepsilon'$ with $E_F$ the Fermi energy. An STM in constant current mode follows a surface of constant current, which translates into a surface of constant integrated LDOS
[$\bar{\rho}(x,y,z,\varepsilon)=C$ with $C$ a real constant]. For each $C$, this construction returns a height $z$ as a function of the position $(x,y)$. This heightmap is then mapped linearly onto a corresponding color scale.\\
\textit{Experiment.} The presented studies were performed in different experimental stations (MAX-lab and BESSY) under identical experimental conditions allowing for a reproducible sample quality in different experiments. As a substrate, the same W(110) single crystal was used in all experiments. Prior to every preparation cycle, a well-established cleaning procedure of the tungsten substrate was applied: several cycles of oxygen treatment with subsequent flashes to $2300^\circ$C. A well-ordered smooth Ni(111) film with a thickness of more than $250$\,\AA\ was prepared by thermal evaporation of Ni metal onto a clean W(110) substrate and subsequent annealing at $300^\circ$C. An ordered graphene overlayer was prepared via thermal decomposition of propene gas (C$_3$H$_6$) according to the recipe described elsewhere~\cite{Dedkov:2010a,Dedkov:2008b,Weser:2011,Dedkov:2008a,Weser:2010}. Deposition of Al was performed at room temperature by thermal evaporation from an Al ingot placed in a cold-lip Knudsen cell crucible and its thickness was controlled by a quartz microbalance as well as by the attenuation of the C\,$1s$ photoemission peak. Intercalation of aluminum was achieved by short annealing at $400^\circ$\,C. The quality, homogeneity, and cleanliness of the prepared systems was verified by means of LEED and core-level as well as valence-band photoemission. The base pressure in all experiments was below $7\times10^{-11}$\,mbar.

NEXAFS studies were performed at the D1011 beamline of the MAX-lab Synchrotron Facility (Lund, Sweden) at both C\,$K$ and Ni\,$L_{2,3}$ absorption edges in partial (repulsive potential $U=-100$\,V) and total electron yield modes (PEY and TEY, respectively) with an energy resolution of 80\,meV and 200\,meV, respectively. All absorption measurements were performed at 300K.

ARPES experiments were performed at the UE56/2-PGM-2 beamline at BESSY (Berlin, Germany). The experimental station consists of two chambers: preparation and analysis. The sample preparation procedure [oxygen-treatments and flashing of W(110) as well as the preparations of the graphene-based systems] was performed in the preparation chamber after which sample was transferred into the analysis chamber for further photoemission measurements. The photoemission intensity data sets $I(E_{kin},k_x,k_y)$ were collected with a PHOIBOS100 energy analyzer (SPECS) while the sample was placed on a 6-axis manipulator. The temperature of the sample during the measurements was kept at 80\,K or 300\,K. The energy/angular resolution was set to 80\,meV/$0.2^\circ$. 

\section{Results and discussion}\label{results}

\begin{figure}
\center
\includegraphics[width=\textwidth]{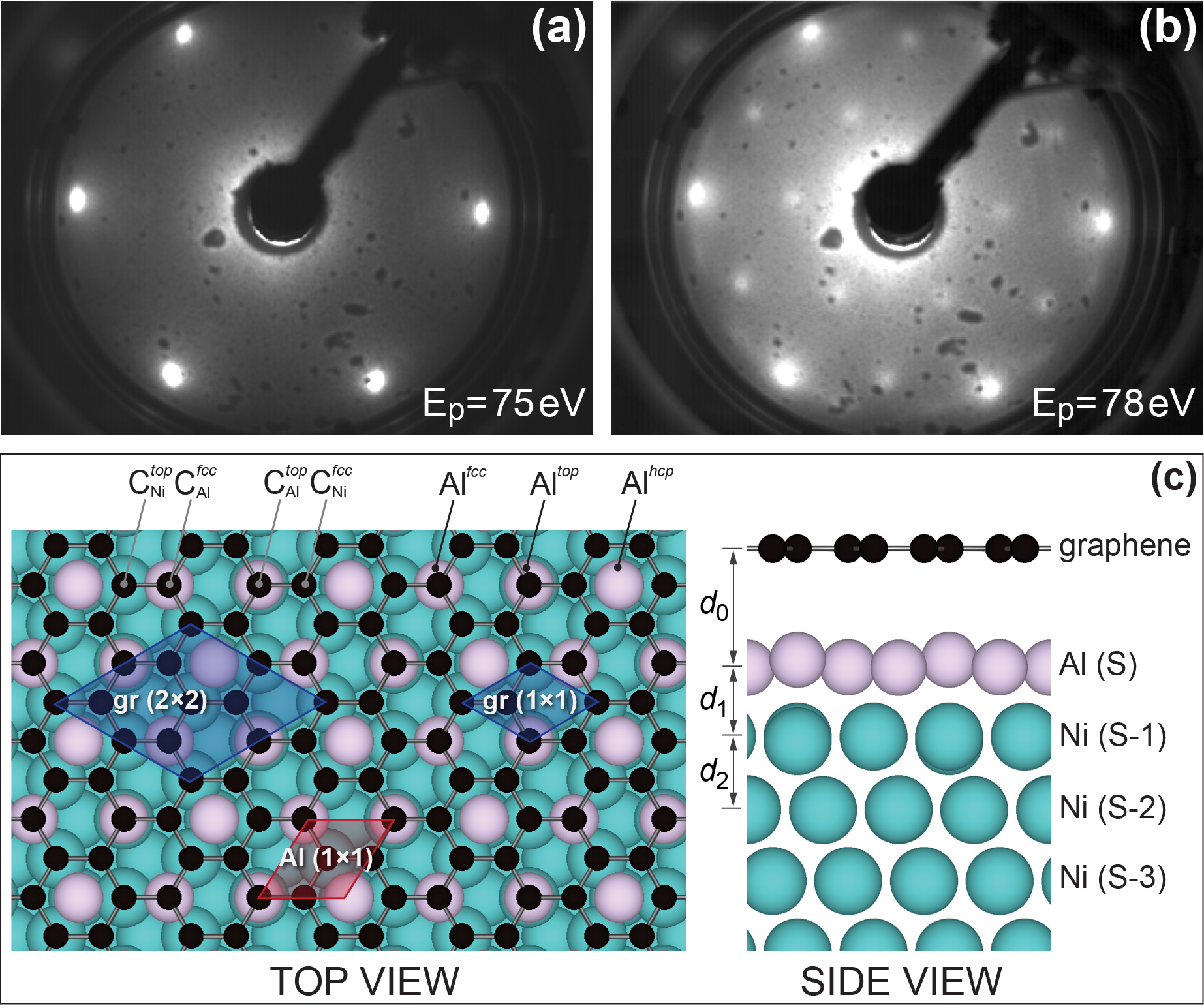}
\caption{LEED images of (a) graphene/Ni(111) and (b) the graphene/Al/Ni(111) system collected at a primary electron energy of 75\,eV and 78\,eV, respectively. (c) Top and side view of the ball model of graphene/Al/Ni(111) obtained after structural optimization. The corresponding graphene and aluminum unit cells as well as C and Al atoms in different occupation positions are marked in the figure.}
\label{LEEDstructure}
\end{figure}

\begin{figure}
\center
\includegraphics[scale=0.7]{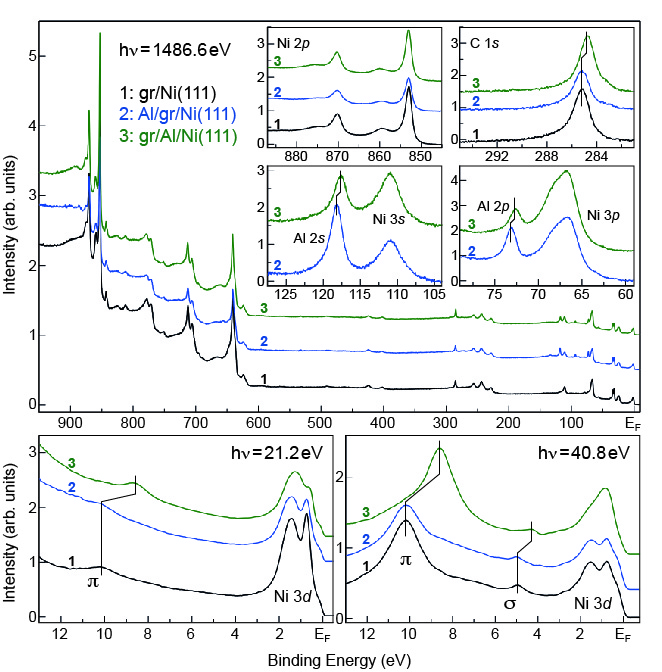}
\caption{(Upper panel) X-ray photoemission spectra (overview as well as for the particular core levels) and (lower panels) normal emission valence band photoemission spectra obtained during preparation of the graphene/Al/Ni(111) system. Corresponding preparation steps, photon energies, core-level peaks, and valence band emission lines are labled in the respective parts of the figure.}
\label{xps}
\end{figure}

The process of formation of the graphene/Al/Ni(111) intercalation-like system was controlled by LEED and core-level as well as valence-band PES in normal emission geometry. These results are compiled in Figures~\ref{LEEDstructure}(a,b) and \ref{xps}, respectively. A ball model of the graphene/Al/Ni(111) system obtained after structural optimization (see discussion below) is shown in Fig.~\ref{LEEDstructure}(c) (top and side view). The unit cells of the graphene layer and the Ni(111) surface have a very small lattice mismatch of about 1\% leading to a hexagonal picture with $p(1\times1)$ symmetry in LEED [Fig.\ref{LEEDstructure}(a)] without any visible additional reflexes indicating the high quality of the graphene/Ni(111) system. Intercalation of 1\,ML Al (ML\,=\,monolayer) underneath graphene on Ni(111) leads to drastic changes in the LEED picture [Fig.\ref{LEEDstructure}(b)]. Firstly, a clear $(2\times2)$ overstructure with respect to the one for the graphene/Ni(111) is clearly visible in LEED indicating formation of the new intercalation-like system with different symmetry. Additionally, weak ring-shaped reflexes ($\approx15^\circ$ away with respect to the main diffraction spots) are visible around the original $(1\times1)$ spots from the graphene layer on Ni(111). A similar ring-shaped structure was observed earlier in Ref.~\cite{Shikin:1998,Dedkov:2001} for the graphene/Cu/Ni(111) system and was assigned to the existence of a misfit between the graphene layer and the underlying Cu layer, which results from a weak graphene-Cu bonding. 

The widely accepted in literature the structure of graphene/Ni(111) has the carbon atoms arranged in the so-called $top-fcc$ configuration on Ni(111)~\cite{Karpan:2007,Karpan:2008,Weser:2011,Bertoni:2004,Gamo:1997}: one carbon atom from the graphene unit cell is placed above Ni interface atom and the second one occupies the hollow $fcc$ site of the Ni(111) surface [marked as ``gr$(1\times1)$'' in Fig.~\ref{LEEDstructure}(c)]. Our present and recent~\cite{Weser:2011} calculations support this model.

\begin{table}
\caption{Results for the atomic structure of the three graphene/metal interface models and for the clean metal surfaces: $E_\mathrm{ads}$(meV) is the adsorption energy per graphene unit cell ``gr$(1\times1)$''; $d_0$(\AA) is the mean distance between the graphene overlayer and the interface metal layer; $d_1$(\AA) is the mean distance between the interface metal layer and the second metal layer; $d_2$(\AA) is the mean distance between the second and third metal layers;  $m_\mathrm{FM}$($\mu_B$) is the interface/surface FM spin magnetic moment (the two values for the $sp$ (upper value) and $d$ (lower value) magnetizations are indicated); $m_\mathrm{C}$($\mu_B$) is the interface carbon spin magnetic moment (the two values for the two nonequivalent carbon atoms are indicated).}
\begin{indented}
\item[]\begin{tabular}{@{}lllllll}
\br
 &Ni(111)&gr/Ni(111)&Al(111)&gr/Al(111)&Al/Ni(111)&gr/Al/Ni(111)\\
\mr
$E_\mathrm{ads}$& &$-321$& &$-104$& &$-87$ \\
 & & & & & & \\
$d_0$& &$2.089$& &$3.188$& &$3.312$\\
 & & & & & & \\                                  
$d_1$&$2.005$&$2.020$&$2.343$&$2.331$&$2.116$&$2.081$\\              
 & & & & & & \\
$d_2$&$2.031$&$2.017$&$2.318$&$2.321$&$2.057$&$2.055$\\
 & & & & & & \\
$m_\mathrm{FM}$&$-0.030$&$-0.024$& & &$-0.021$&$-0.017$\\
                                                &$0.710$&$0.528$& & &$0.407$&$0.273$\\
 & & & & & & \\                                                
$m_\mathrm{C}$& &$-0.019$& & $0.000$& &$0.000$\\
                                             & &$0.033$& & & & \\
\br
\end{tabular}
\end{indented}
\label{structure_table}
\end{table}

In the graphene/Al/Ni(111) system the $fcc$ Al(111) monolayer is placed in the space between graphene and Ni(111) [Fig.~\ref{LEEDstructure}(c)]. There is only one possible arrangement of aluminum atoms in the high-symmetry positions in such a system as shown in the figure: the Al(111) lattice plane is rotated by $30^\circ$ with respect to the graphene/N(111) lattice such that the Al atoms occupy all three different high-symmetry adsorption sites in the space between the graphene layer and the Ni(111) surface. They are noted in Fig.~\ref{LEEDstructure}(c) with respect to the adsorption sites of the Ni(111) surface when occupying $fcc$ (Al$^{fcc}$), $hcp$ (Al$^{hcp}$), and $top$ (Al$^{top}$) positions. In this case the Al layer has a structure $(2\sqrt{3}/3\times2\sqrt{3}/3)R30^\circ$ with respect to graphene/Ni(111) [marked as ``Al$(1\times1)$'' in Fig.~\ref{LEEDstructure}(c)]. The resulting supercell for the graphene/Al/Ni(111) system is marked as ``gr$(2\times2)$'' in Fig.~\ref{LEEDstructure}(c). In this structure there are four different occupation sites for carbon atoms in the unit cell with respect to the adsorption sites of the Ni(111) surface when they are placed above a Ni or Al atom. They are labled in Fig.~\ref{LEEDstructure}(c) as C$^{top}_\mathrm{Ni}$, C$^{fcc}_\mathrm{Ni}$, C$^{top}_\mathrm{Al}$, and C$^{fcc}_\mathrm{Al}$, respectively. The distance between Al atoms in this structure is $2.877$\,\AA, which is very close to the one of $2.863$\,\AA\ extracted from the Al $fcc$ bulk structure.

The obtained crystallographic model of the graphene/Al/Ni(111) intercalation-like system was used for the calculation its electronic structure (see also discussion below). As a first step, as described in Sec.~\ref{experimental}, this structure was allowed to relax with respect to distances between atomic layers (the $z$ coordinates of the graphene layer, Al layer, and two top Ni layers were relaxed) and then the electronic and magnetic properties were calculated for the obtained atoms arrangement. The resulting structure is shown as a side view in Fig.~\ref{LEEDstructure}(c) and the corresponding mean distances between layers are presented in Tab.~\ref{structure_table}.

The graphene layer on top of Al/Ni(111) is extremely flat with mean distance between graphene and Al layer of $3.312$\,\AA\ and the maximal variation in height of about $0.007$\,\AA. This result is very close to the one recently published for the graphene/Au/Ni(111) system~\cite{Kang:2010}. The mean distance between the underlying Al layer and Ni(111) is $2.081$\,\AA. This Al layer is strongly buckled and variations of the distance between graphene and the Al layer are $-0.146$\,\AA, $+0.078$\,\AA, and $+0.068$\,\AA\ for the Al atoms in $top$, $hcp$, and $fcc$ positions with respect to the Ni(111) surface. The mean distance between top Ni (S-1) and Ni (S-2) layers is $2.055$\,\AA\ and the top Ni layer is also buckled with a maximal variation in height of about $0.131$\,\AA. Deeper in the bulk the Ni slab adapts the distance between layers characteristic for bulk Ni of $2.028$\,\AA.

Figure~\ref{xps} shows core level (upper panel) and valence band photoemission spectra (lower panels) recorded during preparation steps of the graphene/Al/Ni(111) system. The corresponding photon energies and the main photoemission features (core level peaks as well as valence band emission lines) are marked in the figure. The photoemission spectra (shape of core level and valence band spectra as well as positions of main photoemission features) of the graphene/Ni(111) system are in very good agreement with the previously obtained results~\cite{Dedkov:2008,Dedkov:2010a} (spectra 1 in Fig.~\ref{xps}). The binding energies (BEs) of the C\,$1s$ peak is $285.1$\,eV and the bottom of the graphene $\pi$ band is at $10.15$\,eV. Deposition of $1$\,ML Al on graphene/Ni(111) leads to the decreasing intensity of photoemission features from graphene and Ni without any visible energy shifts of these lines (spectra 2 in Fig.~\ref{xps}). The formation of the graphene/Al/Ni(111) system [short annealing of Al/graphene/Ni(111) at $400^\circ$\,C] leads to drastic changes in photoemission spectra (spectra 3 in Fig.~\ref{xps}). Firstly, the intensities of the C\,$1s$ peak and the graphene $\pi$ states are restored. At the same time the intensities of Al\,$2s, 2p$ peaks are reduced. These facts indicate that graphene intercalates underneath graphene on Ni(111). After intercalation, the C\,$1s$ peak and the graphene $\pi$ states are shifted by $\approx0.45$\,eV and $\approx1.6$\,eV to lower BEs, respectively,  with regard to those for the graphene/Ni(111) system indicating the electronic decoupling of the graphene layer from the substrate. The Al layer, which is placed after intercalation between graphene and Ni(111), is characterized by Al\,$2s, 2p$ photoemission peaks shifted to lower binding energies [with respect to energies for the Al/graphene/Ni(111) system] indicating the interaction between Al and Ni and a partial charge transfer in the system. These observations are consistent with the previously reported data for the Ni-Al systems~\cite{Chang:2000,Ohtsu:2007}. The corresponding changes are also visible in the region of the Ni\,$3d$ states in the vicinity of $E_F$.

\begin{figure}
\center
\includegraphics[scale=0.62]{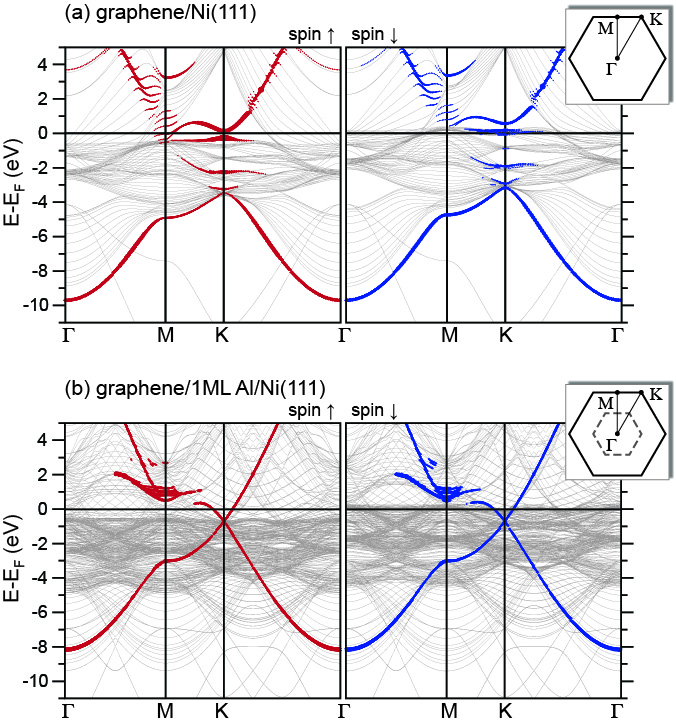}
\caption{Spin-resolved electronic band structure of (a) graphene/Ni(111) and (b) graphene/Al/Ni(111) presented along the $\Gamma-\mathrm{M}-\mathrm{K}-\Gamma$ way in the hexagonal Brillouin zone of graphene. The corresponding Brillouin zones for the ``gr$(1\times1)$'' and ``gr$(2\times2)$'' unit cells are shown in insets by solid and dashed lines, respectively. In the case of graphene/Al/Ni(111) the folded bands due to the $(2\times2)$ structure of the system are present in the plot. The weight of the graphene-derived $p_z$ character is highlighted by the size of filled circles superimposed with the plot of the band structure.}
\label{bandstr}
\end{figure}

A comparison of the calculated electronic structures of the graphene/Ni(111) and the graphene/Al/Ni(111) systems is shown in Fig.~\ref{bandstr}(a,b) (see also supplementary material: Figs.\,S1--S5). The calculated electronic structure of the graphene layer on Ni(111) is in very good agreement with the previously published results~\cite{Bertoni:2004,Weser:2011} [Fig.~\ref{bandstr}(a)] and it is discussed here only briefly. In this system the original electronic structure of the free-standing graphene is strongly disturbed due to the strong hybridization of the graphene $\pi$ and Ni\,$3d$ states. This interaction leads to the shift of graphene-derived electronic bands to higher BE and the appearance of several interface states in the vicinity of $E_F$ with the predominant contributions of graphene $\pi$ and Ni\,$3d_{z^2}$ character~\cite{Bertoni:2004,Weser:2011,Dedkov:2010a}. As a consequence, the magnetic moment of the interface Ni atoms is reduced and an induced magnetic moment on the carbon atoms appears in the graphene layer (Tab.~\ref{structure_table}) that was confirmed experimentally~\cite{Weser:2010,Weser:2011,Dedkov:2010a}.

The electronic band structure of the graphene/Al/Ni(111) system is different from the one of graphene/Ni(111) [Fig.~\ref{bandstr}(b)]. Insertion of the Al layer between a graphene layer and Ni(111) decouples the electronic structure of graphene from the substrate, preventing thereby hybridization between graphene $\pi$ and Ni\,$3d$ valence band states. The Dirac point in the graphene/Al/Ni(111) system is fully restored, but it is shifted by $0.686$\,eV below $E_F$ indicating electron doping of graphene on Al/Ni(111). This value is very close to $0.57$\,eV calculated recently for graphene/Al(111) by means of the local-density approximation with a distance of $3.41$\,\AA\ between a graphene layer and the Al(111) surface~\cite{Khomyakov:2009}. The main $\pi$ branches are clearly recognizable in the electronic structure of graphene/Al/Ni(111) and they almost reproduce the electronic structure of the free-standing graphene~\cite{Bertoni:2004,Weser:2011} (except for a downwards shift). Compared to graphene/Ni(111), in the present system there is no strong hybridization between graphene and Al valence band states preventing Dirac cone. There is only one energy region (away from $E_F$) where such hybridization is visible: around $\mathrm{M}$ at approximately $E-E_F=1$\,eV  (Fig.~\ref{bandstr} and see also partial bands structures in the supplementary material). Also in this region one can clearly see the disparity between different carbon atoms in the ``gr$(2\times2)$'' unit cell as the hybridization of the graphene $\pi$ states from different carbon atoms with the Al\,$3p$ and Ni\,$3d$ valence band states leads to a splitting of the graphene-derived $\pi$ band along $\mathrm{M}$ to $\mathrm{K}$. Here one can conclude that the symmetry in the graphene lattice is broken. Surprisingly, such difference in the symmetry of carbon atoms in the graphene layer in the graphene/Al/Ni(111) system does not lead to any sizable energy gap for the $\pi$ states around the $\mathrm{K}$ point of Brillouin zone (see zoom around the $\mathrm{K}$ point in Fig.\,S6 of the supplementary material). The appearance of such gap was recently demonstrated by ARPES for the graphene layer on Cu(111), Ag(111), and Au(111)~\cite{Enderlein:2010,Varykhalov:2010a}. However, these observations are not supported by band structure calculations where very small~\cite{Kang:2010} or no energy gap was observed for these surfaces~\cite{Khomyakov:2009}. 

\begin{figure}
\center
\includegraphics[scale=0.45]{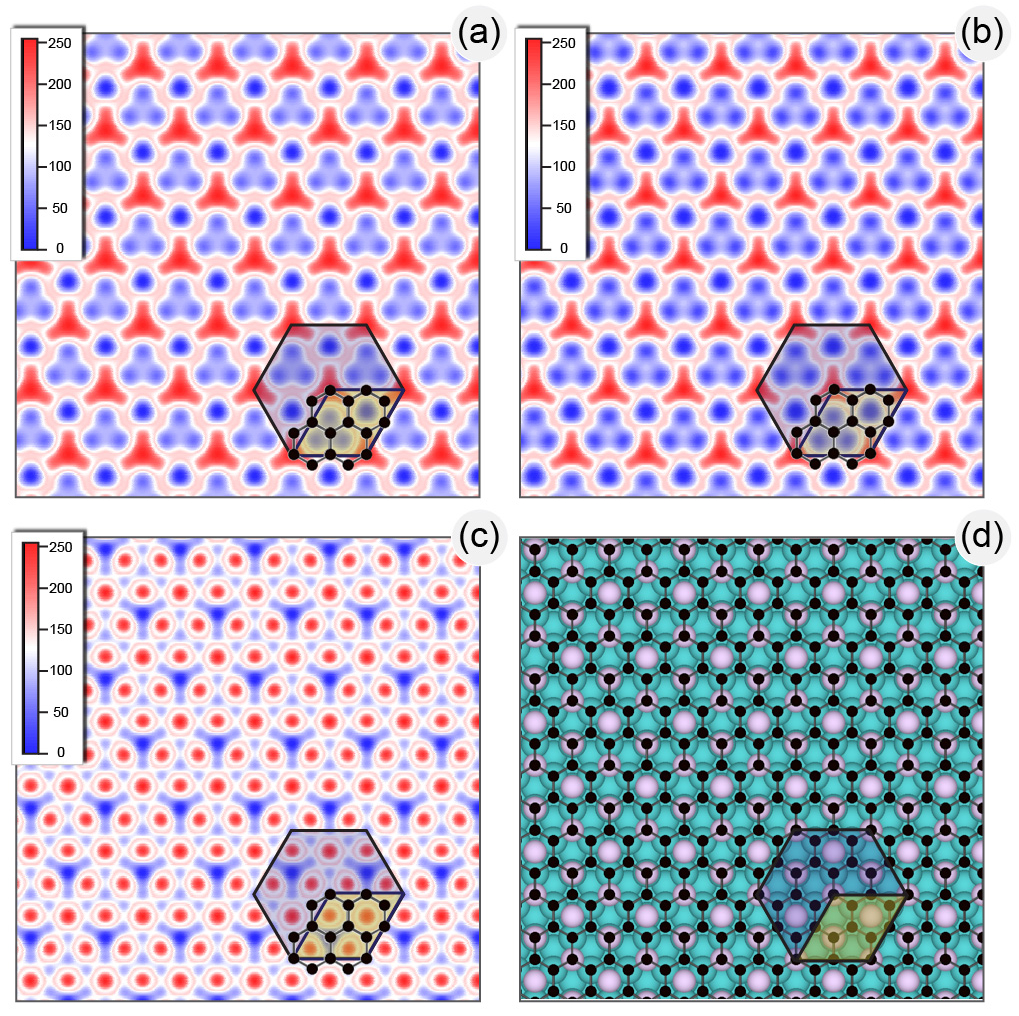}
\caption{(a-c) Calculated STM images for the graphene/Al/Ni(111) system for the distance between the tip and the surface of $3.5$\,\AA. For these images the integrations were performed for the valence band states in the energy range (a) $E-E_F=-2...0$\,eV, (b) $E-E_F=-1...0$\,eV, and (c) $E-E_F=0...2$\,eV that corresponds to the the surfaces of constant density of $0.012\,e/\mathrm{\AA}^3$, $0.005\,e/\mathrm{\AA}^3$, and $0.026\,e/\mathrm{\AA}^3$, respectively. The corresponding region of the lattice is shown in (d). All images are superimposed with a hexagon for the ``gr$(2\times2)$'' unit cell.}
\label{stm_calc}
\end{figure}

In order to get information about the spacial distribution of the valence band states in the vicinity of $E_F$ the STM images of the graphene/Al/Ni(111) were calculated (Fig.~\ref{stm_calc} and Fig.\,S7 of supplementary material), for occupied (a,b) and unoccupied (c) states, respectively. All images are superimposed on the ``gr$(2\times2)$'' unit cell. First of all these results demonstrate that the image contrast for most places in the STM picture is changed upon changing the sign of sample bias voltage. This can be connected with an increase of density of states of Al\,$3s, 3p$ character for unoccupied states (see supplementary material for the atom projected band structures). The interesting fact is that the space symmetry is clearly broken in the graphene unit cell, which would lead, as discussed earlier, to the appearance of an energy gap for the graphene $\pi$ states at the $\mathrm{K}$ point. However, as was shown earlier no such gap is obtained in the calculated band structure. The only gap for the $\pi$ states is visible along the $\mathrm{M}-\mathrm{K}$ way in the Brillouin zone that could be connected with the broken symmetry. The obtained results have to be proved by comparison to scanning tunneling microscopy/spectroscopy measurement for graphene/Al/Ni(111), and with an analysis of the symmetry of the electronic states at the surface in order to get more insight on the origin of the energy gap in the electronic structure of graphene on different substrates. Results similar the ones presented here were recently obtained for the graphene/Au/Ni(111) system where an analysis of calculated band structure and calculated STM was performed~\cite{Kang:2010}.

The present results of band structure calculations of the modification of the electronic structure of graphene on Ni(111) upon intercalation of 1\,ML Al are now compared with experimental spectroscopic data for unoccupied and occupied valance band states obtained by means of NEXAFS and ARPES, respectively.

\begin{figure}
\center
\includegraphics[scale=0.45]{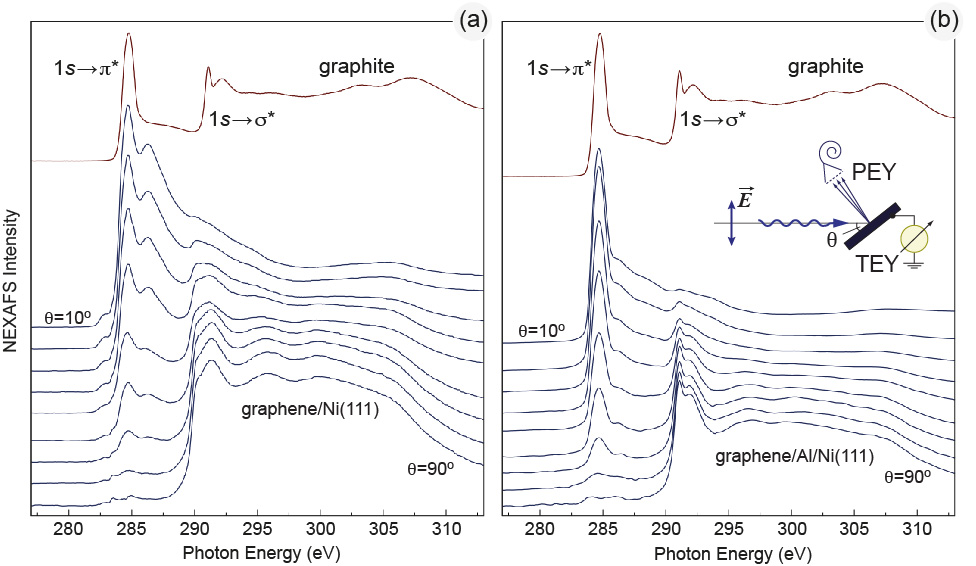}
\caption{Series of angle-dependent NEXAFS spectra of (a) graphene on Ni(111) and (b) the graphene/Al/Ni(111) intercalation-like system recorded at the C\,$K$ absorption edge with an increasing angle $\theta$ from top to bottom [see inset in (b)]. For comparison the NEXAFS spectrum of bulk graphite measured at $\theta=30^\circ$ is shown in the upper part of every panel.}
\label{nexafsC1s}
\end{figure}

\begin{figure}
\center
\includegraphics[scale=0.5]{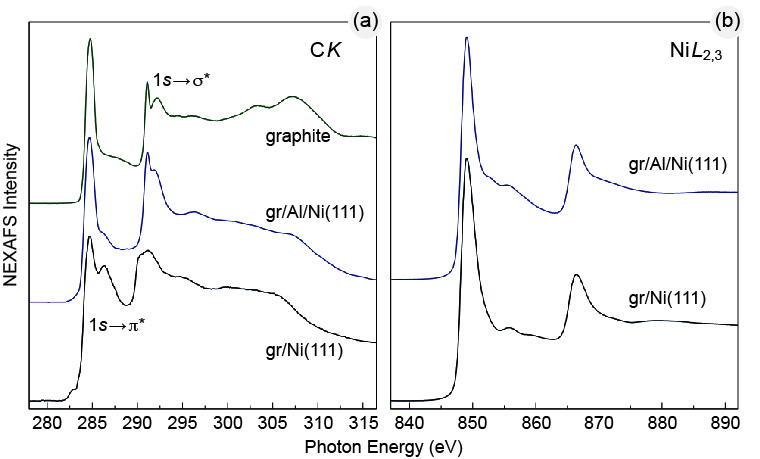}
\caption{Comparison of NEXAFS spectra recorded with $\theta=40^\circ$ at (a) the C\,$K$ and (b) Ni\,$L_{2,3}$ absorption edges for graphite, graphene/Ni(111), and graphene/Al/Ni(111).}
\label{nexafs_compare}
\end{figure}

The unoccupied valence band states of graphene/Ni(111) and graphene/Al/Ni(111) were studied by means of NEXAFS spectroscopy at the C\,$K$ and Ni\,$L_{2,3}$ absorption edges (Figs.~\ref{nexafsC1s} and ~\ref{nexafs_compare}). Angle-dependent NEXAFS spectra shown in Fig.~\ref{nexafsC1s} for both systems represent a nice example demonstrating the so-called search-light-like effect, which in general can be used for probing the quadrupole moment of the local charge around the absorbing atom. In such an experiment, the absorption intensity associated with a specific molecular orbital final state has a maximum if the electric field vector is aligned parallel to the direction of maximum charge or hole density, i.\,e. along a molecular orbital, and the intensity vanishes if the electric field vector is perpendicular to the orbital axis. These spectra show an increased intensity for the $1s\rightarrow\pi^*$ transition in case of grazing incidence of light, and higher intensity for the $1s\rightarrow\sigma^*$ transition in case of normal incidence of light on the sample surface.

The C\,$K$ NEXAFS spectra of the graphene/Ni(111) system [Fig.~\ref{nexafsC1s}(a)] have been studied and discussed in detail in Refs~\cite{Weser:2010,Dedkov:2010a,Rusz:2010}. In the region of the $1s\rightarrow\pi^*$ transition these spectra have a double-peak structure compared to those of graphite that is explained by the transitions of C\,$1s$ core electron into two unoccupied states (interface states) which are the result of hybridization of the graphene $\pi$ and Ni valence band states: the first feature corresponds to the transition into the interface state above the Fermi level (around the $\mathrm{K}$-point in the Brillouin zone) which originates from C\,$p_z$--Ni\,$3d$ hybridization and corresponds to carbon atom C-$top$ and interface Ni atom antibonding state; the second peak in the spectrum corresponds to a transition of the electron into the interface state above the Fermi level (around the $\mathrm{M}$-point in the Brillouin zone) which originates from C\,$p_z$--Ni\,$p_x,p_y,3d$ hybridization and corresponds to a bonding between the two carbon atoms, C-$top$ and C-$fcc$, which involves the nickel interface atom (see Fig.~\ref{bandstr}(a) and discussion in Refs.~\cite{Weser:2010,Dedkov:2010a,Rusz:2010,Weser:2011}). The visible reduction in the energy separation between the $\pi^*$ and $\sigma^*$ features for all spectra compared to that in spectra of graphite is a result of the lateral bond softening within the adsorbed graphene monolayer.

Intercalation of thin Al layer underneath a graphene layer on Ni(111) leads to drastic changes in the C\,$K$ NEXAFS spectra [Fig.~\ref{nexafsC1s}(b) and Fig.~\ref{nexafs_compare}(b)]. The shape of these spectra, positions of main spectroscopic features  as well as the energy separation between $\pi^*$ and $\sigma^*$ features become similar to those in the spectra of pure graphite. These facts immediately indicate that the graphene layer is decoupled from the substrate by intercalated Al. These observations are also consistent with the main conclusions of the band structure calculations. The intense main $\pi^*$ peak in the NEXAFS spectra of graphene/Al/Ni(111) can be assigned to the transition of a $1s$ core electron into the unoccupied states around the $\mathrm{M}$-point at $\approx0.8$\,eV above $E_F$ [Fig.~\ref{bandstr}(b)]. The second spectroscopic feature at $286.3$\,eV of photon energy (small shoulder in the region of the $1s\rightarrow\pi^*$ resonance) can be assigned to the transition of the electron from the $1s$ level on the graphene unoccupied states around the $\mathrm{M}$-point at $\approx2.8$\,eV above $E_F$ [Fig.~\ref{bandstr}(b)].

\begin{figure}
\center
\includegraphics[scale=0.5]{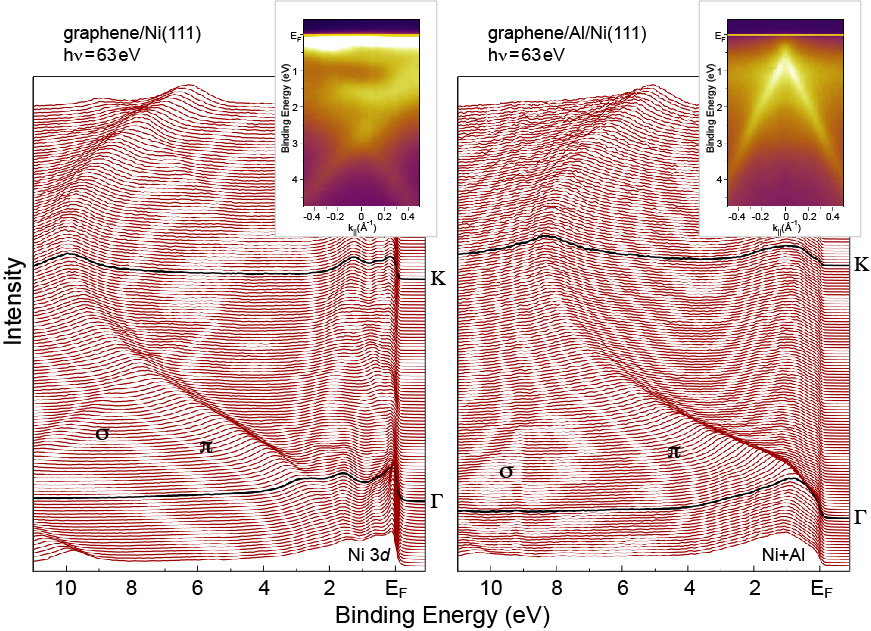}
\caption{Series of angle-resolved photoemission spectra measured along the $\Gamma-\mathrm{K}$ direction in the Brillouin zone for (a) graphene/Ni(111) and (b) the graphene/Al/Ni(111) system. Insets show the photoemission intensity maps around the $\mathrm{K}$-point along the direction perpendicular to $\Gamma-\mathrm{K}$. All data sets were obtained with $h\nu=63$\,eV photon energy.}
\label{arpes}
\end{figure}

Figure~\ref{nexafs_compare}(b) shows Ni\,$L_{2,3}$ NEXAFS spectra of graphene/Ni(111) and graphene/Al/ Ni(111) obtained in PEY mode (repulsive potential $U=-100$\,V) allowing to perform measurements in a more surface/interface-sensitive regime. A noticeable modification of this spectrum after intercalation of the Al layer can be clearly visible: the shape and energy of the so-called $6$\,eV Ni correlation satellite are modified; additionally a small shoulder at $\approx852.6$\,eV of photon energy is visible. The resulting Ni\,$L_{2,3}$ NEXAFS spectrum is similar, to some extent, to that of Ni$_3$Al~\cite{Chang:2000} and these spectral changes are explained by the charge transfer from Al on Ni. These experimental observations of the interaction at the Al/Ni interface are supported by the band structure calculations [Fig.~\ref{bandstr}(b), Tab.~\ref{structure_table}]. This charge transfer leads to a strong reduction of the Ni\,$3d$ magnetic moment from $0.528\,\mu_B$ for graphene/Ni(111) to $0.273\,\mu_B$ in graphene/Al/Ni(111). As a consequence, the interaction between the graphene layer and the substrate is reduced and graphene becomes freestanding-like with a small electron doping that shifts the Dirac cone below $E_F$.

The electronic structure of occupied states below $E_F$ of graphene/Ni(111) and graphene/Al/Ni(111) was studied by means of ARPES and results of these experiments are shown in Fig.~\ref{arpes}(a,b), respectively. Here the angle-resolved photoemission data along the $\Gamma-\mathrm{K}$ direction and around the $\mathrm{K}$-point along the direction perpendicular to $\Gamma-\mathrm{K}$ of the Brillouin zone are presented. The experimental data shown in Fig.~\ref{arpes}(a) for the graphene/Ni(111) system are in very good agreement with the previously published experimental and theoretical data~\cite{Dedkov:2001,Shikin:2000,Dedkov:2010a,Weser:2011,Bertoni:2004,Gruneis:2008} and demonstrate the strong hybridization of the graphene $\pi$ and Ni\,$3d$ valence band states leading to a partial charge transfer from Ni to graphene and a destruction of the Dirac cone in graphene. This effect is clearly seen in the inset of Fig.~\ref{arpes}(a) where hybrid states ``cross'' at $\approx2.8$\,eV BE, forming a series of interface states in the vicinity of $E_F$. This experimental picture is consistent with the recent theoretical interpretation~\cite{Bertoni:2004,Dedkov:2010a,Weser:2011}.

Interesting changes in photoemission data obtained in experiments appear after intercalation of the Al-layer underneath graphene on Ni(111). The resulting photoemission spectra presented in the same way as for graphene/Ni(111) are shown in Fig.~\ref{arpes}(b). First of all, as was discussed earlier, the intercalation of Al leads to the decoupling of the electronic states of the graphene layer from those of the substrate. This results in the shift of all electronic bands of graphene to the lower binding energies by values presented earlier (see Fig.~\ref{xps} and the corresponding discussion). Secondly, the electronic structure of the graphene layer as well as the Dirac cone in the vicinity of $E_F$ are fully restored. As was discussed earlier and it is supported here experimentally there is a small electron doping of graphene leading to the shift of the Dirac point below $E_F$ by $0.64$\,eV. This value is in very good agreement with the theoretically calculated value of $0.686$\,eV presented earlier in this work [Fig.~\ref{bandstr}(b)]. The small difference between experimental and theoretical data could be assigned to the limitations of the slab calculation.

\begin{figure}
\center
\includegraphics[scale=0.75]{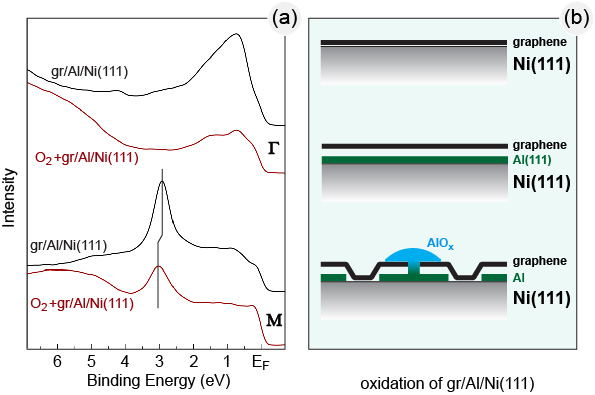}
\caption{(a) Angle-resolved photoemission spectra measured at the $\Gamma$ and $\mathrm{M}$-point of Brillouin zone for graphene/Al/Ni(111) before and after an attempt to perform an oxidation of the Al-layer underneath graphene. The clear restoring of Ni\,$3d$ emission and the shift of the graphene $\pi$ states to higher binding energies are visible. (b) The proposed model for the mechanism of the oxidation of the graphene/Al/Ni(111) system.}
\label{grAlNi-ox}
\end{figure}

In our experiments we undertake an attempt to form a fully epitaxial graphene/AlO$_x$/Ni(111) system on the basis of our Al-intercalation-like sandwich. The precondition for this experiment is the recently published works on the intercalation of oxygen underneath graphene on Ru(0001)~\cite{Zhang:2009a,Sutter:2010a}. Here one can hope that the same mechanism can be applied for the oxidation of the thin Al-layer underneath graphene on Ni(111). This would allow to prepare the system which could be used in future devices where a graphene layer is bottom-gated and where spin injection can be effectively performed without considerable spin scattering at the interfaces.

The main results of these experiments are shown in Fig.~\ref{grAlNi-ox}(a) (see also Fig.\,S8 in supplementary material for discussion of the XPS results) and the model for the oxidation of the graphene/Al/Ni(111) system is presented in Fig.~\ref{grAlNi-ox}(b). First of all, the room-temperature oxidation of graphene/Al/Ni(111) [$T=300$\,K, $p(\mathrm{O}_2)=1\times10^{-5}$\,mbar, 30\,min] was not found to lead to any changes in photoemission spectra and LEED images of the system, again proving the complete intercalation of the Al-layer underneath graphene on Ni(111) (XPS and ARPES spectra are the same as shown in Fig.~\ref{xps}). Further oxidation at slightly higher temperature and for longer time [$T=450$\,K, $p(\mathrm{O}_2)=1\times10^{-5}$\,mbar, 60\,min] leads to strong changes in the spectra. XPS spectra (see Supplementary material) of the system after this step show the appearance of an AlO$_x$ component on the Al\,$2s,2p$ emission lines, indicating partial oxidation of the thin Al layer; at the same time the intensity of the C\,$1s$ emission line is decreased. In the valence band photoemission spectra shown in Fig.~\ref{grAlNi-ox}(a) one can clearly see the O\,$2p$ emission below $4$\,eV BE, partial restoration of the Ni\,$3d$ emission in the vicinity of $E_F$, and a shift of the graphene $\pi$ emission line to slightly higher binding energies by $\approx120$\,meV (these states are also broadened compared to the former case). LEED images after the second oxidation step become very diffuse. The observed changes in photoemission spectra and LEED images can be explained by the model shown in Fig.~\ref{grAlNi-ox}(b). After the first oxidation step (room temperature oxidation) the graphene layer on Al/Ni(111) behaves like an protective layer as demonstrated earlier in Refs.~\cite{Dedkov:2008,Dedkov:2008b}. Increasing the temperature during oxidation leads to the interaction of carbon atoms at defects of the graphene layer with oxygen, leading to an increasing of number of such defects as well as their effective area. This leads to the interaction of non-protected aluminium with oxygen and the formation of an AlO$_x$ layer on top of such defects [see Fig.~\ref{grAlNi-ox}(b)]. Here, the relatively strong bonds between Ni and Al at the interface are broken and this process is analogous to aluminothermic reaction~\cite{Wang:1993}. A similar effect was recently observed in Refs.~\cite{Dedkov:2002b,Dedkov:2007b} for the oxidation of the Al/Fe(110) interface.

\section{Conclusions}

In the present work, the electronic structure of the Al-intercalation-like system on the basis of graphene/Ni(111) is studied via a combination of structural (LEED) and spectroscopic methods (NEXAFS, XPS, ARPES) accompanied by state-of-the-art electronic structure calculations. A thin Al layer leads to the decoupling of the graphene layer from the substrate, such that it behaves like a quasi-freestanding film with a small electron doping, leading to the shift of the Dirac cone to approximately $0.64$\,eV below $E_F$. The experimental results are in very good agreement with theoretical predictions. Intercalation of Al can help to overcome some difficulties which may appear in the analogous systems on the basis of $d$-electron noble-metals where spin scattering in the graphene-based spin filters may show up. Unfortunately, the approach used in the present work to prepare a fully epitaxial graphene/AlO$_x$/Ni(111) system leads to the pulling out of Al on top of the system with the subsequent formation of AlO$_x$ layer on top of the graphene/Ni(111) system.

\section*{Acknowledgements}

We would like to thank A. Preobrajenski (MAX-lab, Lund) for his support during beamtime. D.\,E.\,P. Vanpoucke is acknowledged for kindly providing us with the STM visualization software ``Hive''. This work has been supported by the European Science Foundation (ESF) under the EUROCORES Program EuroGRAPHENE (Project ``SpinGraph''). E.\,N.\,V. appreciate the support from DFG through the Collaborative Research Center (SFB) 765 ``Multivalency as chemical organization and action principle: New architectures, functions and applications''.Y.\,S.\,D. acknowledges the financial support by the German Research Foundation under project DE\,1679/2-1. We acknowledge the technical assistance by MAX-lab and BESSY. We appreciate the support from the HLRN (The North-German Supercomputing Alliance).

\section*{References}


\newpage
\noindent
Supplementary material for manuscript:\\
\textbf{Structural and electronic properties of the graphene/Al/Ni(111) intercalation-like system}\\
\newline
E. N. Voloshina,$^{1}$ A. Generalov,$^{2,3}$ M. Weser,$^{2}$ S. B\"ottcher,$^{2}$ K. Horn,$^{2}$ and Yu.\,S.\,Dedkov$^{2,3}$\\
\newline
$^1$\mbox{Physikalische und Theoretische Chemie, Freie Universit\"at Berlin, 14195 Berlin, Germany}\\
$^2$Fritz-Haber-Institut der Max-Planck-Gesellschaft, 14195 Berlin, Germany\\
$^3$Institut f\"ur Festk\"orperphysik, Technische Universit\"at Dresden, 01062 Dresden, Germany\\
\newline
\textbf{List of figures:}
\\
\noindent\textbf{Fig.\,S1.} Detailed analysis of the electronic structure of the graphene/1\,ML\,Al(111)/ Ni(111) system where the corresponding weights of  C-$2p_z$-projected bands are shown by thick lines.

\noindent\textbf{Fig.\,S2.} Detailed analysis of the electronic structure of the graphene/1\,ML\,Al(111)/ Ni(111) system where the corresponding weights of  Al-$3s$-projected bands are shown by thick lines.

\noindent\textbf{Fig.\,S3.} Detailed analysis of the electronic structure of the graphene/1\,ML\,Al(111)/ Ni(111) system where the corresponding weights of  Al-$3p$-projected bands are shown by thick lines.

\noindent\textbf{Fig.\,S4.} Detailed analysis of the electronic structure of the graphene/1\,ML\,Al(111)/ Ni(111) system where the corresponding weights of  Ni-$3d$-projected bands are shown by thick lines.

\noindent\textbf{Fig.\,S5.} Detailed analysis of the electronic structure of the graphene/1\,ML\,Al(111)/Ni(111) system where the corresponding weights of  Ni-$3d_{z^2}$-projected bands are shown by thick lines.

\noindent\textbf{Fig.\,S6.} Zoom of the electronic structure of he graphene/1\,ML\,Al(111)/ Ni(111) system around the $\mathrm{K}$ point. 

\noindent\textbf{Fig.\,S7.} Calculated STM images for the graphene/Al/Ni(111) system for the distance between the tip and the surface of $3.0$\,\AA, $3.5$\,\AA, and $4.0$\,\AA\ (vertical columns). For these images the integrations were performed for the valence band states in the energy range $E-E_F=-2...0$\,eV, $E-E_F=-1...0$\,eV, and $E-E_F=0...2$\,eV (horizontal rows). The corresponding constant densities are marked in each image. All images are superimposed with the mark for the ``gr$(2\times2)$'' unit cell, similar to Fig.\,4 of the main manuscript.

\noindent\textbf{Fig.\,S8.} Oxidation of the graphene/Al/Ni(111) system in conditions marked in the figure.

\clearpage
\begin{figure}
\includegraphics[width=0.65\textwidth]{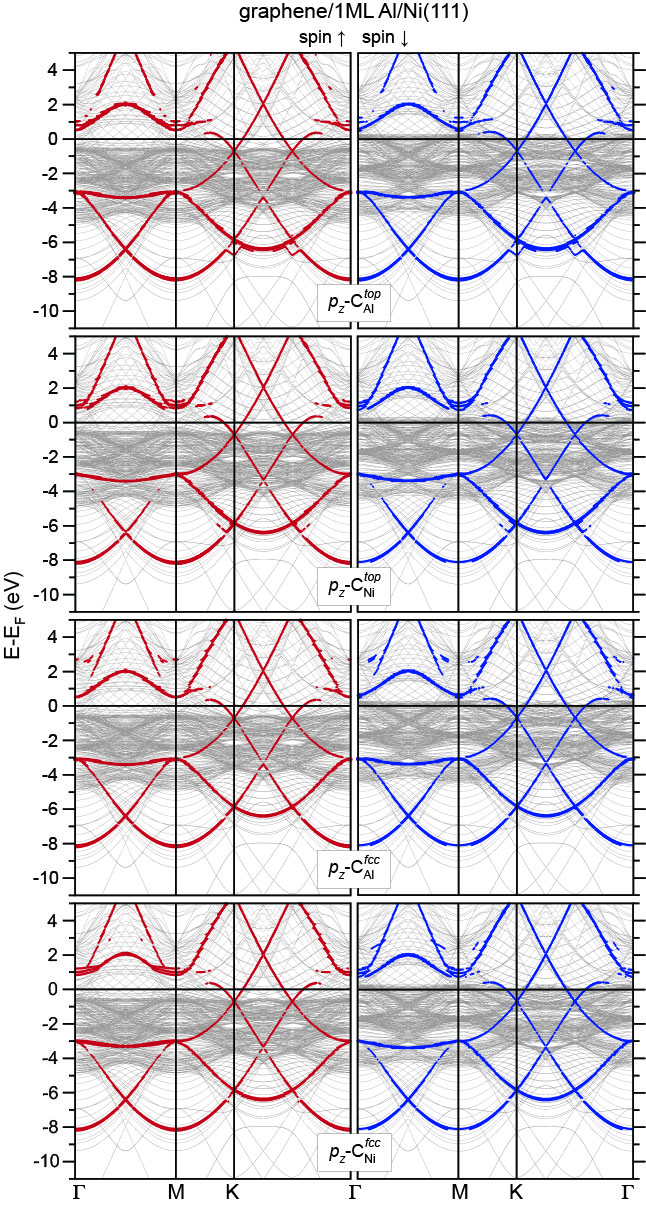}
\end{figure}
\noindent\textbf{Fig.\,S1.} Detailed analysis of the electronic structure of the graphene/1\,ML\,Al(111)/ Ni(111) system where the corresponding weights of  C-$2p_z$-projected bands are shown by thick lines.

\clearpage
\begin{figure}
\includegraphics[width=0.65\textwidth]{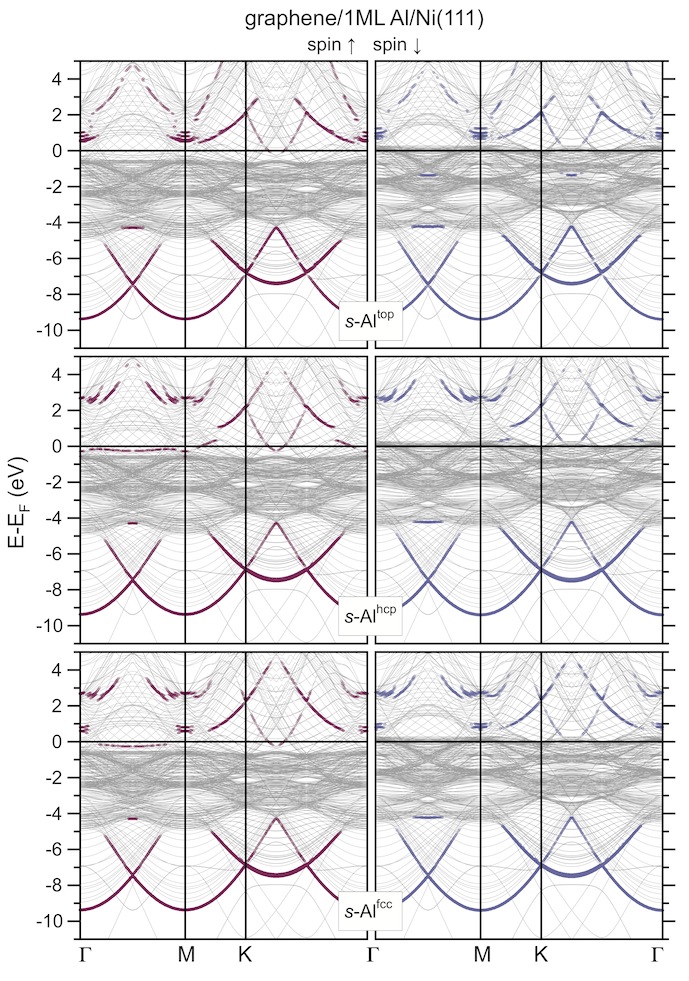}
\end{figure}
\noindent\textbf{Fig.\,S2.} Detailed analysis of the electronic structure of the graphene/1\,ML\,Al(111)/ Ni(111) system where the corresponding weights of  Al-$3s$-projected bands are shown by thick lines.

\clearpage
\begin{figure}
\includegraphics[width=0.65\textwidth]{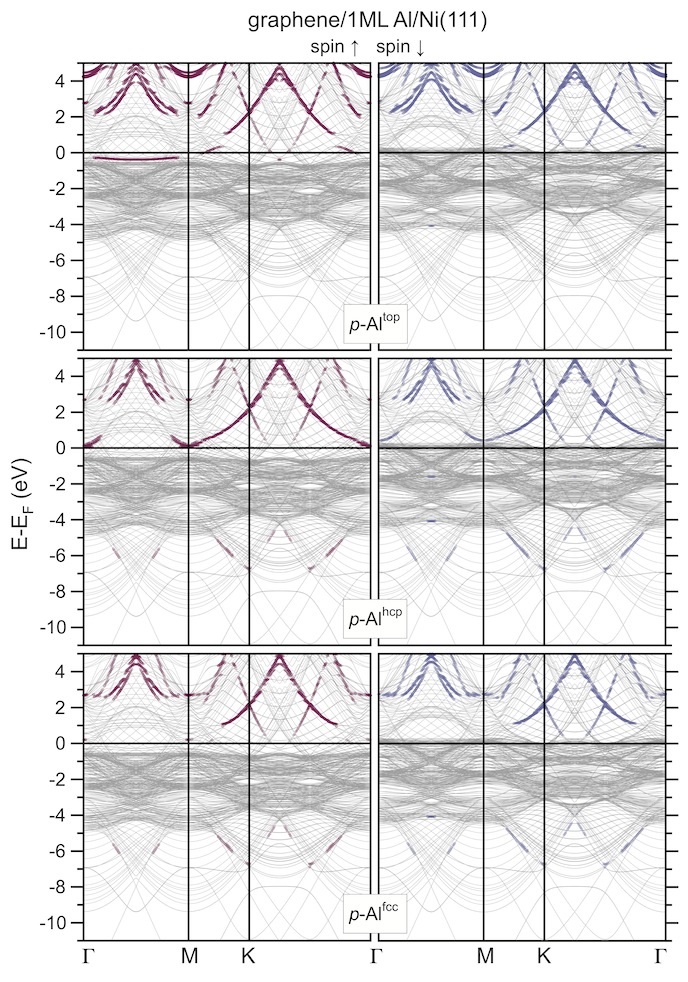}
\end{figure}
\noindent\textbf{Fig.\,S3.} Detailed analysis of the electronic structure of the graphene/1\,ML\,Al(111)/ Ni(111) system where the corresponding weights of  Al-$3p$-projected bands are shown by thick lines.

\clearpage
\begin{figure}
\includegraphics[width=0.85\textwidth]{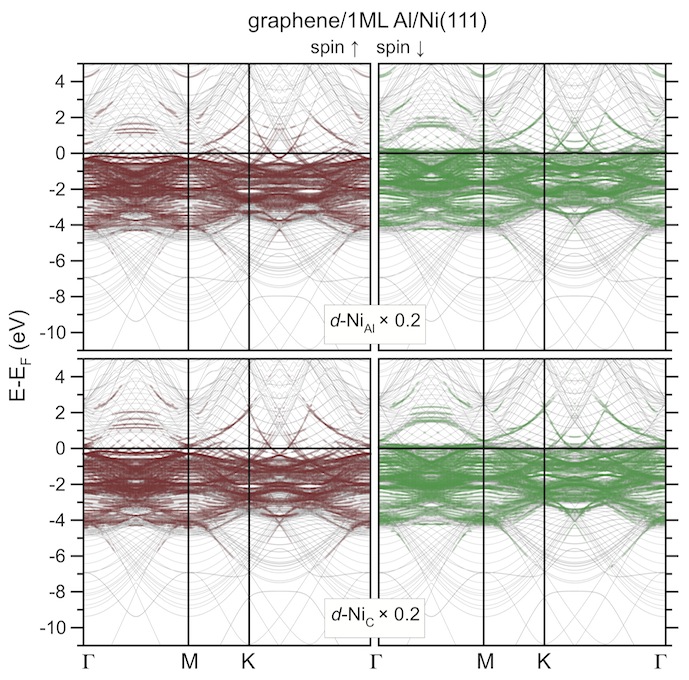}
\end{figure}
\noindent\textbf{Fig.\,S4.} Detailed analysis of the electronic structure of the graphene/1\,ML\,Al(111)/ Ni(111) system where the corresponding weights of  Ni-$3d$-projected bands are shown by thick lines.

\clearpage
\begin{figure}
\includegraphics[width=0.85\textwidth]{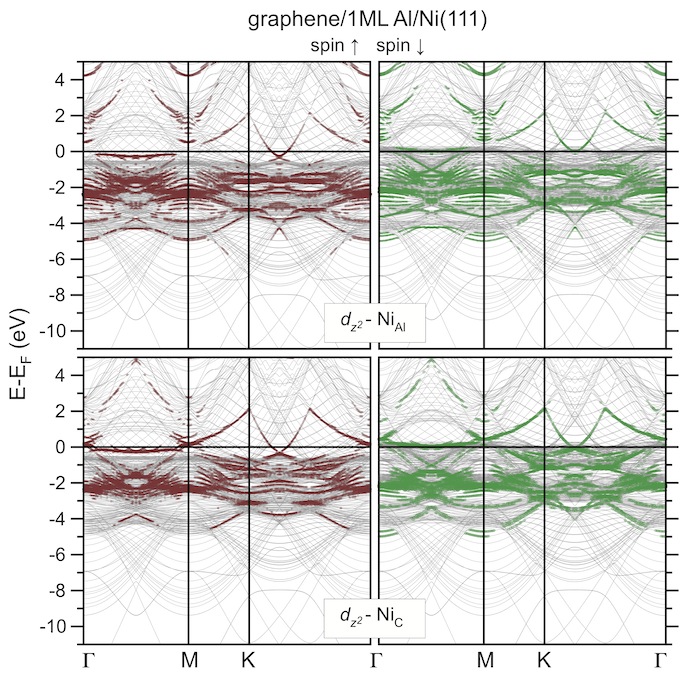}
\end{figure}
\noindent\textbf{Fig.\,S5.} Detailed analysis of the electronic structure of the graphene/1\,ML\,Al(111)/ Ni(111) system where the corresponding weights of  Ni-$3d_{z^2}$-projected bands are shown by thick lines.

\clearpage
\begin{figure}
\includegraphics[width=0.45\textwidth]{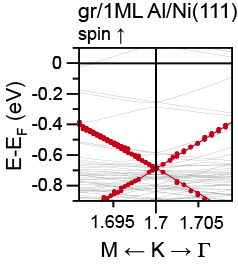}
\end{figure}
\noindent\textbf{Fig.\,S6.} Zoom of the electronic structure of he graphene/1\,ML\,Al(111)/ Ni(111) system around the $\mathrm{K}$ point. 

\clearpage
\begin{figure}
\includegraphics[scale=0.5]{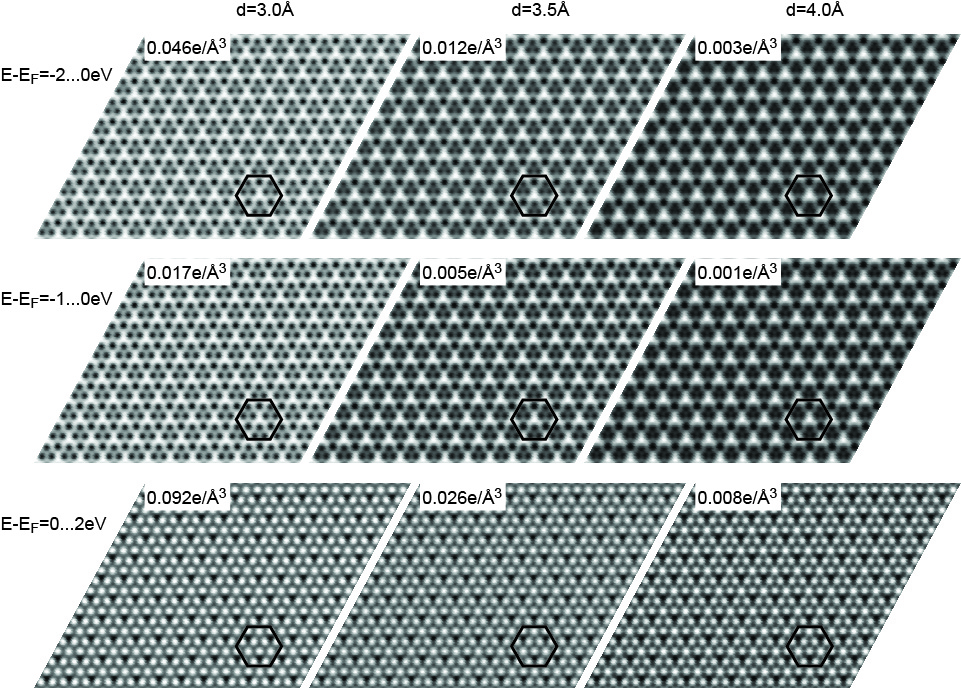}
\end{figure}
\noindent\textbf{Fig.\,S7.} Calculated STM images for the graphene/Al/Ni(111) system for the distance between the tip and the surface of $3.0$\,\AA, $3.5$\,\AA, and $4.0$\,\AA\ (vertical columns). For these images the integrations were performed for the valence band states in the energy range $E-E_F=-2...0$\,eV, $E-E_F=-1...0$\,eV, and $E-E_F=0...2$\,eV (horizontal rows). The corresponding constant densities are marked in each image. All images are superimposed with the mark for the ``gr$(2\times2)$'' unit cell, similar to Fig.\,4 of the main manuscript.

\clearpage
\begin{figure}
\includegraphics[scale=0.65]{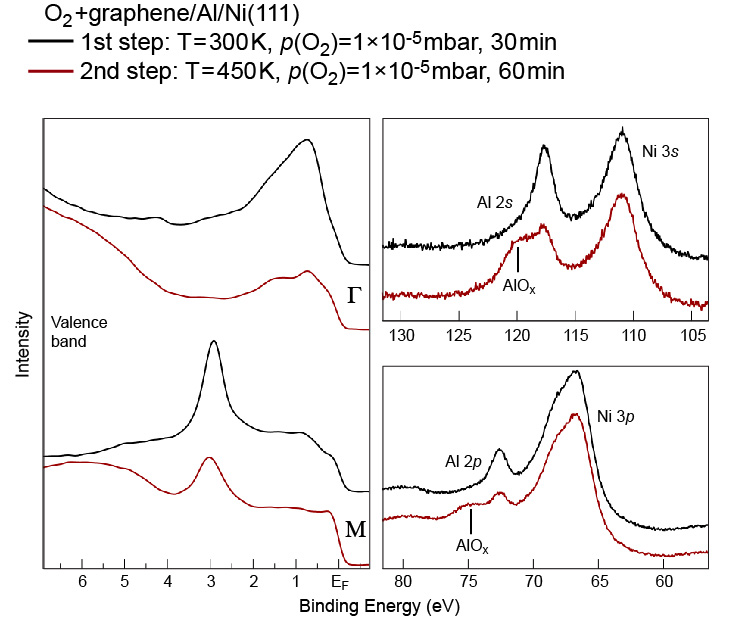}
\end{figure}
\noindent\textbf{Fig.\,S8.} Oxidation of the graphene/Al/Ni(111) system in conditions marked in the figure. The corresponding marks for valence band region and corresponding core level are placed in the figure.

\end{document}